\documentclass[aps,prl,reprint,amsmath,amssymb,superscriptaddress]{revtex4-1}

\usepackage[colorlinks,citecolor=blue,linkcolor=blue]{hyperref}
\usepackage{graphicx}
\usepackage{dcolumn}
\usepackage{bm}
\usepackage{color}

\newcommand{\CNN}{Centre de Nanosciences et de Nanotechnologies, CNRS, Univ. Paris-Sud, Universit\'e Paris-Saclay, 91120 Palaiseau, France}
\newcommand{\UMPhy}{Unit\'e Mixte de Physique, CNRS, Thales, Univ. Paris-Sud, Universit\'e Paris-Saclay, 91767 Palaiseau, France}
\newcommand{\IJL}{Institut Jean Lamour, CNRS, Universit\'e de Lorraine, 54011 Nancy, France}
\newcommand{\LMOPSa}{Chaire Photonique, CentraleSup\'elec, Universit\'e Paris-Saclay, 57070 Metz, France}
\newcommand{\LMOPSb}{Laboratoire Mat{\'e}riaux Optiques, Photonique et Syst{\`e}mes, CentraleSup\'elec, Universit\'e de Lorraine, 57070 Metz, France}

\begin{document}

%
%
\title{Chaos in Magnetic Nanocontact Vortex Oscillators}

\author{Thibaut Devolder}
\affiliation{\CNN}
\author{Damien Rontani}
\affiliation{\LMOPSa}
\affiliation{\LMOPSb}
\author{S{\'e}bastien Petit-Watelot}
\affiliation{\IJL}
\author{Karim Bouzehouane}
\affiliation{\UMPhy}
\author{St{\'e}phane Andrieu}
\affiliation{\IJL}
\author{J{\'e}r{\'e}my L{\'e}tang}
\author{Myoung-Woo Yoo}
\author{Jean-Paul Adam}
\author{Claude Chappert}
\affiliation{\CNN}
\author{St{\'e}phanie Girod}
\author{Vincent Cros}
\affiliation{\UMPhy}
\author{Marc Sciamanna}
\affiliation{\LMOPSa}
\affiliation{\LMOPSb}
\author{Joo-Von Kim}
\email{joo-von.kim@c2n.upsaclay.fr}
\affiliation{\CNN}

\date{1 October 2019}

\begin{abstract}
We present an experimental study of spin-torque driven vortex self-oscillations in magnetic nanocontacts. We find that above a certain threshold in applied currents, the vortex gyration around the nanocontact is modulated by relaxation oscillations, which involve periodic reversals of the vortex core. This modulation leads to the appearance of commensurate but also more interestingly here, incommensurate states, which are characterized by devil's staircases in the modulation frequency. We use frequency- and time-domain measurements together with advanced time-series analyses to provide experimental evidence of chaos in incommensurate states of vortex oscillations, in agreement with theoretical predictions.
\end{abstract}

\maketitle

Chaos describes a deterministic nonlinear dynamical process that is exponentially sensitive to initial conditions. In the context of physical systems such as microelectronic or photonic devices, chaotic behavior has been studied for different possible applications in information technologies~\cite{Ditto:2015fx, Sciamanna:2015ec}, where the underlying premise is that the complexity of a chaotic signal can be harnessed to compute or process information. For example, the high information entropy content of a chaotic signal can be used for random number generation at GHz rates and beyond~\cite{Uchida:2008gx, Li:2013il, Virte:2014bw, Zhang:2017fl}, its symbolic dynamics can be used to encode information~\cite{Hayes:1993ee, Bollt:1997jk, Schweizer:2001uw, Bollt:2003gm}, and the possibly large fractal dimension combined with synchronization capabilities makes it an ideal source for secure communications at the physical level~\cite{Cuomo:1993ba, Argyris:2005br}.

In this context, nanoscale spintronic devices such as spin-torque nano-oscillators~\cite{Silva:2008hi, Slavin:2009fx, Kim:2012du, Chen:2016bf} are promising for chaos-based applications for a number of reasons. First, magnetization dynamics is inherently nonlinear as a result of magnetocrystalline anisotropies, dipolar interactions, and certain nonconservative processes. Second, spin-dependent transport effects, such as spin transfer torques~\cite{Ralph:2008kj}, which allow magnetization dynamics to be driven by electrical currents, and magnetoresistance, which allows such dynamics to be detected electrically, offer promising avenues for integration into micro- and nanoelectronics. In these systems, chaos can appear as a result of periodic driving~\cite{Li:2006gl, Pylypovskyi:2013ce}, as delayed-feedback effects~\cite{Williame:2019hd}, in the dynamics of coupled vortices~\cite{Bondarenko:2019fm}, and during magnetization reversal~\cite{Montoya:2019fj}.

The nanocontact vortex oscillator~\cite{Pufall:2007jc, Mistral:2008js, Ruotolo:2009ex, Kuepferling:2010jn, PetitWatelot:2012be, Keatley:2016et, Keatley:2016ku, Keatley:2017fb} represents an intriguing example, where different commensurate and incommensurate states appear due to competing self-oscillations~\cite{PetitWatelot:2012be}. The primary oscillation is driven by spin torques and involves self-sustained vortex gyration around the nanocontact~\cite{Mistral:2008js}, which is accompanied by relaxation oscillations in the form of periodic core reversal above a threshold current. Commensurate states represent self-phase locking between these two modes, where the ratio of the two frequencies is rational, while for incommensurate phases this ratio is irrational. Simulations have suggested that incommensurate phases lead to a chaotic time series, but this had not been observed directly in our earlier experiments.

In this Letter, we present experimental observations of such incommensurate states in a nanocontact vortex oscillator. By using frequency- and time-resolved measurements together with advanced time series analysis of the magnetization dynamics, we show first signatures in the power spectra and autocorrelation function that are consistent with the chaotic behavior predicted in simulations. We further support these findings using the technique of titration of chaos with added noise~\cite{Poon:PNAS:2001}, which reveals a strong level of nonlinearity only in the incommensurate states, consistent with the presence of chaos.

\begin{figure}
\centering\includegraphics[width=8.5cm]{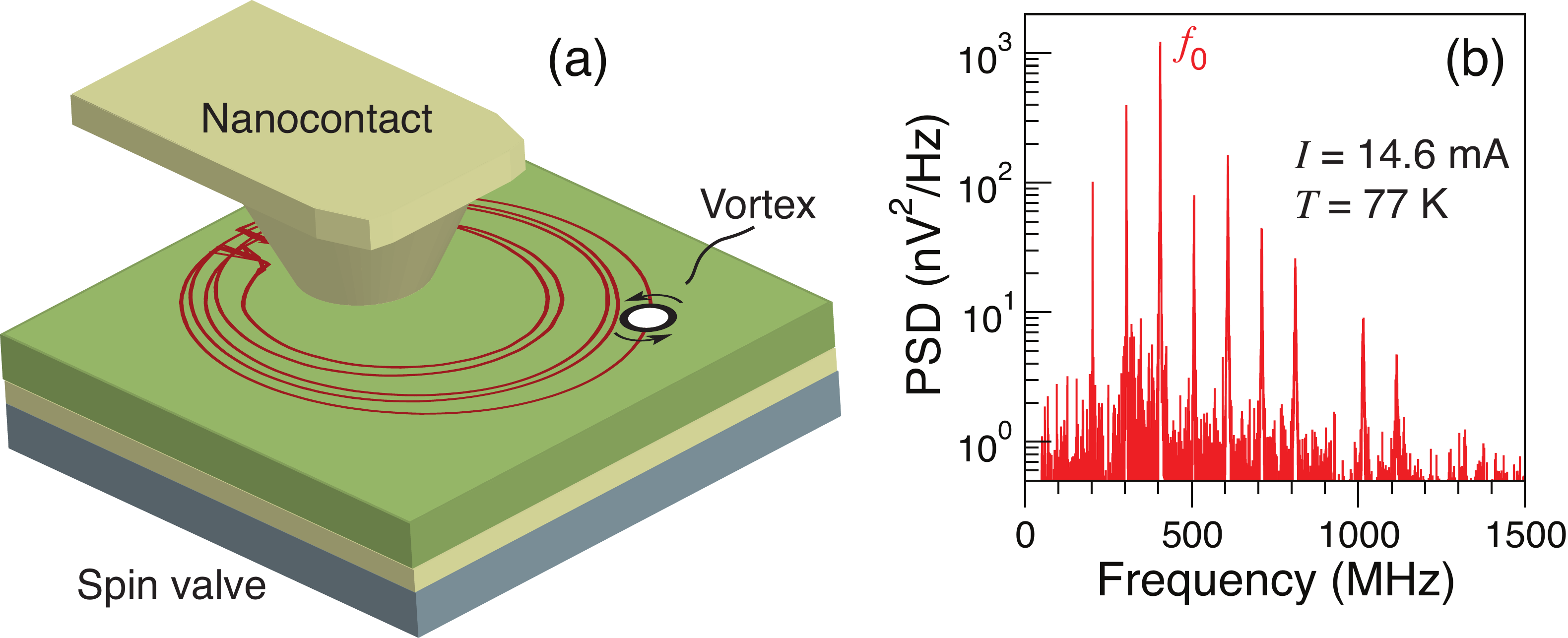}
\caption{(a) Schematic illustration of vortex oscillations in the magnetic nanocontact geometry. The trajectory of the vortex core (red line) illustrates the additional modulation due to vortex core reversals, which take place within a restriction region of the trajectory.  (b) Example of a power spectrum of voltage oscillations showing self-modulation of vortex gyration due to periodic core reversal in the absence of applied magnetic fields.}
\label{fig:intro}
\end{figure}

An illustration of the nanocontact system is presented in Fig.~\ref{fig:intro}(a). The spin valve is an extended multilayered film with the composition SiO$_2$/Cu (40)/Co (20)/Cu (10)/Ni$_{81}$Fe$_{19}$ (20)/Au (6)/photoresist (50)/Au (top contact), where the figures in parentheses are layer thicknesses in nm. The multilayer was grown at room temperature by dc magnetron sputtering in an argon atmosphere with a residual pressure of $6.4 \times 10^{-8}$ mbar. The film was subjected to stabilization annealing during the fabrication process at 170$^\circ$ C for 1 minute. The film magnetic properties were determined prior to patterning using vector network analyzer ferromagnetic resonance. The NiFe layer has the expected soft properties, including a coercivity of 1 mT, a saturation magnetization $\mu_0 M_s = (1.053 \pm 0.003)$ T, a spectroscopic splitting factor of $g = 2.111 \pm 0.003$, and a Gilbert damping constant of $\alpha = (7 \pm 1)\times 10^{-3}$. The Co layer is also relatively soft with a coercivity of 2 mT, with $\mu_0 M_s = (1.768 \pm 0.011)$ T, $g =  2.133 \pm 0.009$, and $\alpha = (10 \pm 1) \times 10^{-3}$. The NiFe layer is the free magnetic layer in which the vortex dynamics takes place, while the Co layer is the reference magnetic layer for the giant magnetoresistance effect. A gold nanocontact of approximately 20 nm in diameter is made on this film using a nanoindentation technique~\cite{Bouzehouane:2003kc}, which involves creating a tapered hole in an ultrathin resist layer using the tip of an atomic force microscope, which allows contact to be subsequently made with the Au layer comprising the top electrode.

The vortex dynamics is initiated and studied as follows. The vortex is first nucleated by reversing the free layer magnetization with an in-plane applied magnetic field in the presence of a static 20 mA current applied through the nanocontact. The Oersted-Ampere field generated by this current~\cite{PetitWatelot:2012iz} leads to the nucleation of a vortex as a domain wall sweeps through the nanocontact area, and the vortex is subsequently confined by the Zeeman potential associated with this field~\cite{Devolder:2011iw}. Spin torques due to the current flowing radially outward from the nanocontact then drive the vortex into a steady state gyration around the nanocontact, which results in magnetoresistance oscillations that are detected after amplification as voltage fluctuations in the frequency domain by a spectrum analyzer and in the time domain by a single-shot oscilloscope. rf switches are used to connect either of these two apparatuses to the sample, hence allowing for both time- and frequency-domain measurements to be made sequentially under the same experimental conditions without switching the dc current off. This precaution is necessary since the induced dynamics is very sensitive to the history of the applied current sweeps, as we discuss below.

The experiments are conducted in a cyrostat at liquid nitrogen temperature to minimize magnetic noise due to thermal fluctuations, which are inherently present due to Joule heating in the nanocontact region that can reach 100 K~\cite{PetitWatelot:2012ik}. This is important for distinguishing between the chaotic processes, which appear as an athermal noise, and stochastic processes that naturally lead to the line shape broadening of the power spectra. An example of the measured power spectrum of vortex oscillations at $T=$ 77 K and in zero applied magnetic field is presented in Fig.~\ref{fig:intro}(b). Under the applied current of 14.6 mA, one observes a spectrum typically associated with the commensurate state in which the central frequency representing the gyration around the nanocontact, $f_0$, appears with a large number of sidebands that result from the additional modulation due to periodic core reversal. In this particular case the modulation frequency is $f_0/4$, giving rise to a phase-locked regime in which the core reversal occurs once after every four revolutions of the vortex core around the nanocontact.

\begin{figure}[b]
\centering\includegraphics[width=8.0cm]{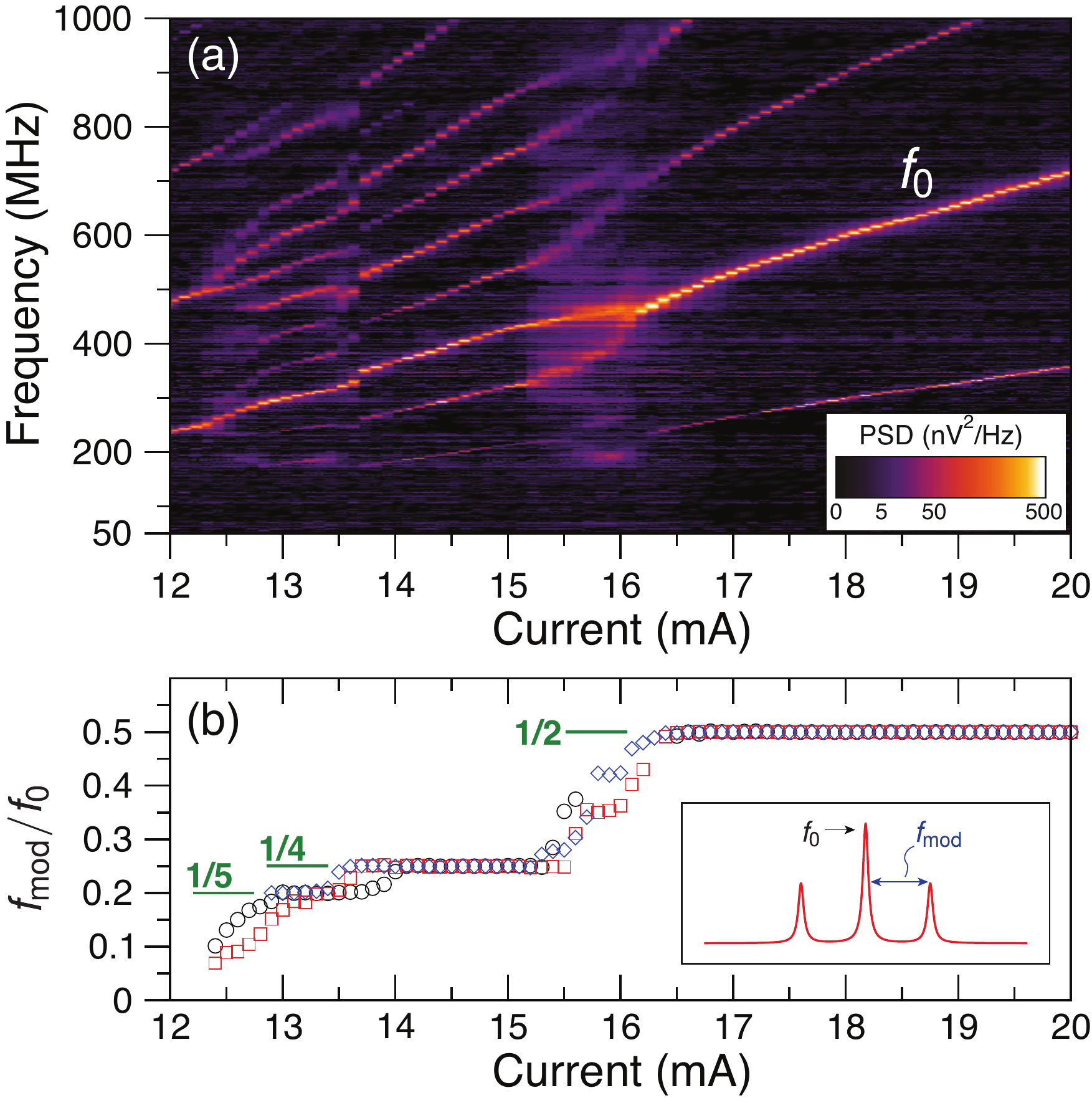}
\caption{(a) Color map of experimental power spectral density as a function of applied current in zero magnetic field. $f_0$ denotes the central frequency. (b) Ratio of the modulation frequency, $f_\mathrm{mod}$, to $f_0$ as a function of applied current. The different colored points correspond to three different current sweeps and the self-phase-locking plateaus are indicated by fractions. The inset shows a schematic of the modulated peak with sidebands.}
\label{fig:PSDvsI}
\end{figure}

The variation of the power spectrum with applied current is presented in Fig.~\ref{fig:PSDvsI}.
Below 12.3 mA, the power spectral density (PSD) of oscillations exhibit no modulation but possesses a rich harmonic content, which is consistent with an elliptical vortex trajectory around the nanocontact~\cite{PetitWatelot:2012be}. This ellipticity can be due to the presence of a remnant antivortex generated from the nucleation process, which remains pinned in close proximity to the nanocontact. As the current is increased above this threshold, modulation sidebands appear as a result of periodic core reversal. Over different current intervals, the ratio between the modulation ($f_{\rm mod}$) and central ($f_{0}$) frequencies are simple integer fractions, as shown in Fig.~\ref{fig:PSDvsI}(b) for the ratios 1/5, 1/4, and 1/2. While these fractions, represented by plateaus in the current dependence of $f_{\rm mod} / f_0$, are reproducible for different current sweeps, the frontier between them are observed to fluctuate between different measurements. This can be seen in Fig.~\ref{fig:PSDvsI}(b) where the results from three different current sweeps are shown. Not only do the positions of the plateaus shift between measurements, but the ratios in the incommensurate states, such as the region between the 1/4 and 1/2 plateaus, also vary from one measured current sweep to the next. We hypothesize that such sensitivity to the history of the current sweeps is related to the position of the remnant antivortex, which has a strong influence on the shape of the vortex trajectory. Nevertheless, these devil's staircases in the modulation frequency exhibit features that are consistent with previous experimental and theoretical results~\cite{PetitWatelot:2012be}.

\begin{figure}
\centering\includegraphics[width=8.5cm]{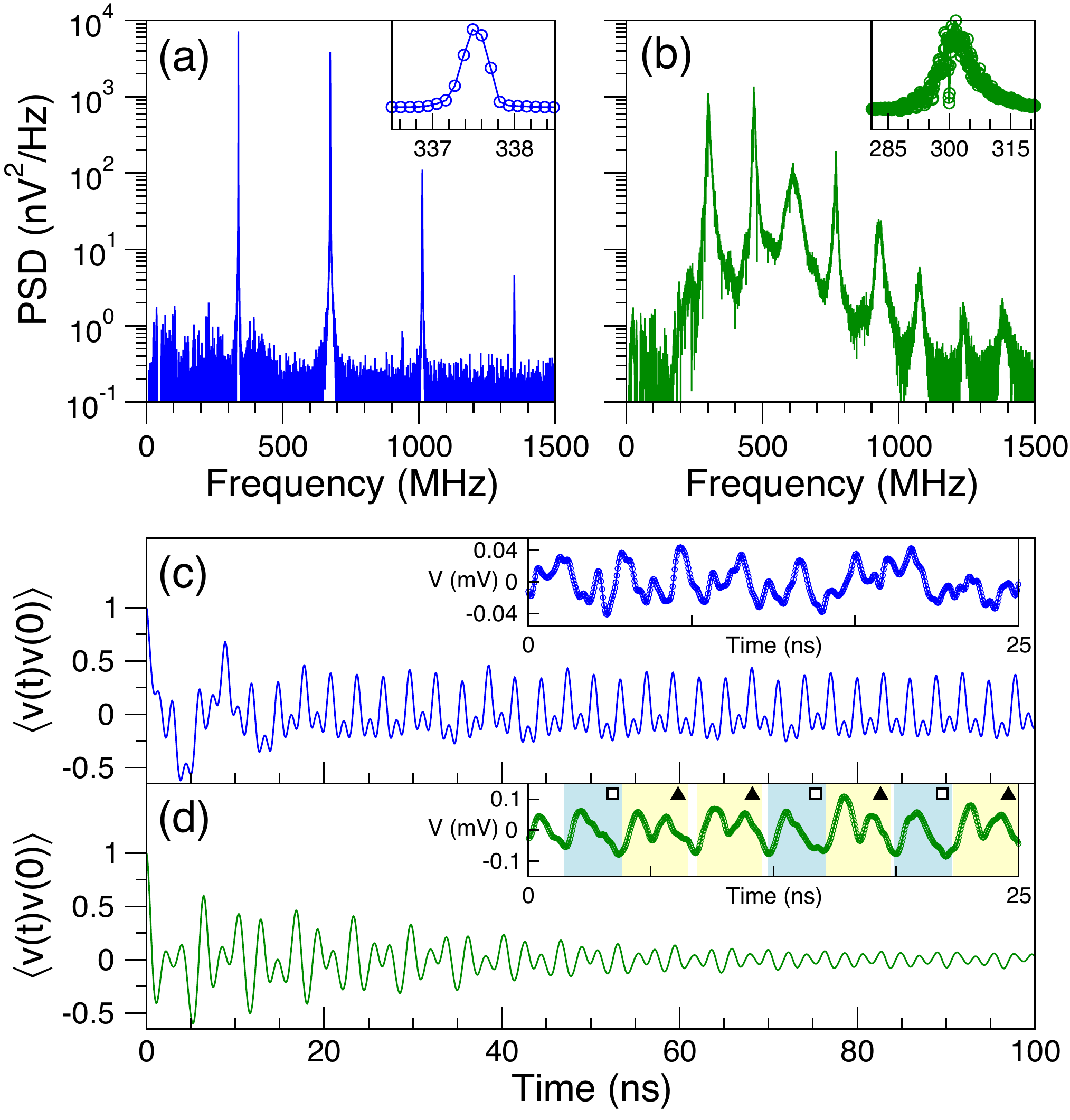}
\caption{Experimental power spectra in the (a) phase-locked state (18 mA, 12.6 mT) and the (b) incommensurate state (18 mA, 4.9 mT). The insets show an enlargement on the lowest frequency peak. Corresponding (normalized) autocorrelation functions of the time traces for (c) the phase locked state and (d) the incommensurate state. The insets show a sample of the single-shot time traces over 25 ns, where in (d) the two identifiable repeating waveforms are denoted by an open square and a filled triangle. }
\label{fig:PSDci}
\end{figure}

We performed time-resolved measurements to investigate the commensurate and incommensurate phases in more detail. We chose to work at a higher current of 18 mA but in the presence of an applied magnetic field oriented perpendicular to the film plane, $H_{\perp}$. The higher current allows for a better signal to noise ratio for the time-domain measurements, while the perpendicular field permits transitions between commensurate and incommensurate phases to occur since it affects in opposite ways the gyration frequencies of vortices of opposite polarities \cite{deLoubens:PRL:2009}. A comparison between the power spectra and time-domain measurements in the commensurate and incommensurate states, obtained at two different applied fields, is presented in Fig.~\ref{fig:PSDci}.
In the frequency domain, the commensurate state is characterized by narrow spectral lines, where the linewidth of the modulation peak is instrument limited and well under 1 MHz [Fig.~\ref{fig:PSDci}(a), inset]. This is smaller than the typical linewidths of $1-3$ MHz for the self-sustained gyration mode without core reversal. This low linewidth indicates that any broadening of the power spectra in this regime is likely to be mainly due to thermal fluctuations. For the incommensurate state, on the other hand, the spectral lines exhibit a significant broadening in addition to the presence of a higher background noise below 1 GHz, as shown in Fig.~\ref{fig:PSDci}(b). The additional noise in this regime is likely to be athermal, since no additional heating of the sample occurs (the current is kept constant) and only the magnetic field strength is varied with respect to the commensurate case.

Since the magnetoresistance signal gives only a projection of the free layer magnetization along the reference layer magnetization, it is difficult to reconstruct the vortex trajectory in the film plane from time-resolved measurements. 
Nevertheless, it is possible to glean some important features from the single-shot time traces and their autocorrelation functions. These are shown in Figs.~\ref{fig:PSDci}(c) and ~\ref{fig:PSDci}(d) for the commensurate and incommensurate states, respectively. In the commensurate phase-locked state, the single-shot traces [Fig.~\ref{fig:PSDci}(c), inset] and their autocorrelation [Fig.~\ref{fig:PSDci}(c)] show a repeating sequence of large and small peaks, which is consistent with a core reversal event occurring after each revolution around the nanocontact~\cite{PetitWatelot:2012be}. The autocorrelation function $\langle v(t) v(0) \rangle$ is normalized. Notice that the decay in the envelope of the oscillations in the autocorrelation function is imperceptible after the initial transient phase of 20 ns, which is consistent with a regime in which the relaxation oscillation is strongly locked to the gyrotropic motion. The situation is qualitatively different in the incommensurate case, where the envelope in $\langle v(t) v(0) \rangle$ decays more rapidly over the same time interval. Nevertheless, there appears to be some correlation in the patterns over the first 30 ns, before being washed out at longer times. These patterns can be seen in the single-shot time traces in the inset of Fig.~\ref{fig:PSDci}(d), where the occurrence of the two identifiable waveform motifs (labeled in the figure by the square and triangle symbols) do not appear to possess any long-time correlations. This behavior is consistent with core reversal events that seem to be randomly distributed in time, which have been shown in zero-temperature simulations to correspond to temporal chaos~\cite{PetitWatelot:2012be}.

\begin{figure}[b]
\centering\includegraphics[width=8.5cm]{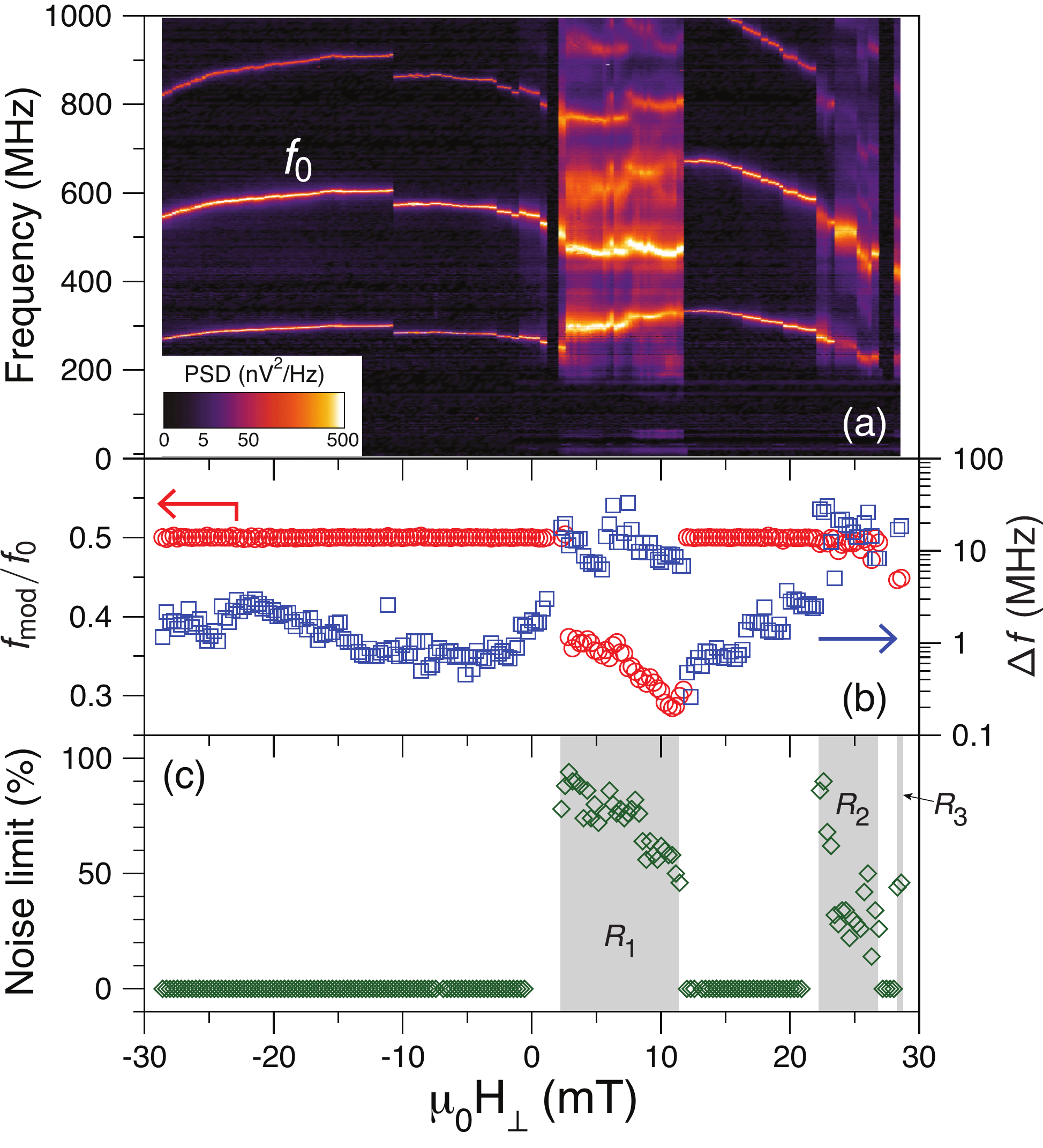}
\caption{(a) Power spectral density map as a function of applied perpendicular field for a fixed current of 18 mA. (b) Ratio of modulation to central frequencies and linewidth of the central peak as a function of applied perpendicular magnetic fields. (c) Noise limit obtained from the noise titration technique, where nonzero values in $R_1$, $R_2$, and $R_3$ are consistent with chaotic dynamics.}
\label{fig:PSDvsH}
\end{figure}

The variation of the PSD with applied perpendicular fields is shown in Fig.~\ref{fig:PSDvsH}. For applied fields $\mu_0 H_\perp < 2.6$ mT, the oscillator remains in a commensurate $f_{\rm mod} / f_0 = 1/2$ state, which is consistent with the behavior presented in Fig.~\ref{fig:PSDvsI}. As the perpendicular field is increased, a transition towards an incommensurate state is observed in which the modulation ratio $f_{\rm mod} / f_0$ takes on a broad range of values from 0.28 to 0.37 in the field range of 2.6 mT $< \mu_0 H_\perp <$ 11.7 mT, as shown in Fig.~\ref{fig:PSDvsH}(b). This transition is accompanied by a large increase in the spectral linewidth of the central peak, $\Delta f$, which is observed to vary by at least an order of magnitude. The linewidth corresponds to the full width at half maximum and is determined from a Lorentzian fit to the $f_0$ peak. For fields above 11.7 mT, the $1/2$ phase-locked state is recovered before another transition to an incommensurate state occurs at 22 mT.

A detailed analysis of the single-shot time series data was performed using the noise titration technique~\cite{Poon:PNAS:2001} to determine whether the complex signal observed arises from a chaotic or stochastic process. Synthetic white noise is added iteratively to the data to reduce progressively the signal-to-noise ratio (SNR) and comparisons are made between one-step prediction errors given by linear and nonlinear models described by discrete Volterra-Wiener series~\cite{Poon:Nature:1996}. When the nonlinear prediction on the degraded-SNR time series is no longer better than the linear prediction, the so-called \textit{noise limit} (NL) is achieved and we stop the titration procedure. The NL follows the same behavior to that of the largest Lyapunov Exponent, traditionally used to assess the presence of chaos~\cite{Kantz:Book:2003}, but is more robust to false-positives in the detection of chaos induced by experimental noise. This is why this method can be preferred for experimental analysis~\cite{Raman:PRL:2006, Poon:PLoS1:2009}. A value of $\text{NL}=0$ is usually a sign of nonchaotic behavior in the data, $0.05<\text{NL}<0.1$ of weak chaos, and $0.1<\text{NL}<1$ of strong chaos; these ranges were obtained with a statistical confidence level of 99$\%$~\cite{Poon:PNAS:2001}.

In Fig.~\ref{fig:PSDvsH}(c), we plot the NL as a function of the transverse magnetic field $\mu_0 H_{\perp}$. We used a second-order discrete nonlinear Volterra-Wiener series with memory depth $\kappa = 15$ and embedding time-delay $\tau_d = 10T_s$ with $T_s = 20\text{ ps}$, the experimental sampling time. The NL is nonzero only in three applied-field regions: $R_1$: $2.4\text{ mT}<\mu_0 H_{\perp}<12\text{ mT}$, $R_2$: $23.4\text{ mT}<\mu_0 H_{\perp}<28.2\text{ mT}$, and $R_3$: $29.7\text{ mT}<\mu_0 H_{\perp}<30\text{ mT}$, which correspond to the regions with incommensurate states and spectral broadening observed in Fig.~\ref{fig:PSDvsH}(b). The range of NL values is $[0.46,0.94]$ for $R_1$, $[0.28,0.9]$ for $R_2$, and $[0.44,0.46]$ for $R_3$, respectively. This is consistent with the presence of a strong level of nonlinearity, and hence chaos (according to the noise titration approach) in the dynamics of vortex self-oscillations, in agreement with theoretical predictions made in Ref.~\cite{PetitWatelot:2012be}.

The capacity to identify chaotic behavior from the time series data from the nanocontact vortex oscillator opens up a number of perspectives for both fundamental and applied studies. The magnetoresistance signal represents an indirect measurement of the vortex core polarity, whose dynamics is challenging to probe electrically. Our study may provide a way of studying the inertial effects and transient dynamics related to core reversal in nanodevices. The chaotic dynamics measured in the magnetoresistance signal is also associated with the erratic generation of regular patterns (as shown in the insert of Fig.~\ref{fig:PSDci}), which could lead to the determination of symbolic dynamics for the system and hence open the way towards controlling the chaotic properties of the oscillator at the nanoscale. Finally, the use of chaotic dynamics in spintronics could lead to the development of novel applications in information processing, such as physical-layer encryption and random number generation~\cite{Fukushima:2014kz}.

\begin{acknowledgments}
The authors thank C. Deranlot for his assistance in sample fabrication. D.R. gratefully thanks C.-S. Poon for sharing information on the implementation of the noise titration technique. This work was supported by the Agence Nationale de la Recherche under Contract No. ANR-17-CE24-0008 (CHIPMuNCS), the Horizon2020 Research Framework Programme of the European Commission under Contract No. 751344 (CHAOSPIN), and the French RENATECH network. The Chaire Photonique is funded by the European Regional Development Fund (ERDF), Ministry of Higher Education and Research (FNADT), Moselle Department, Grand Est Region, Metz Metropole, AIRBUS-GDI Simulation, CentraleSup{\'e}lec, and Fondation Sup{\'e}lec. 
\end{acknowledgments}

\bibliography{articles}
\bibliographystyle{apsrev4-2}

\end{document}